\documentclass[12pt]{elsart}
\usepackage[dvips]{graphicx}
\usepackage{epsfig}
\usepackage{amssymb}

\begin{document}

\begin{frontmatter}

\title{Enhancing the Pierre Auger Observatory to the $10^{17}$ to $10^{18.5}$ eV Range: Capabilities
of an Infill Surface Array}

\author[CNEA]{M.C. Medina\thanksref{cor1}\thanksref{CONICET}},
\author[CAB]{M. G\'{o}mez~Berisso\thanksref{CONI}},
\author[CAB]{I. Allekotte}
\author[CNEA]{A. Etchegoyen\thanksref{CONI}},
\author[MX]{G. Medina Tanco}, and
\author[CNEA]{A.D. Supanitsky},

\address[CNEA]{Departamento de F\'{\i}sica, Comisi\'{o}n Nacional de Energ\'{\i}a
 At\'{o}mica, Av. Gral. Paz 1499, (1650) Buenos Aires, Argentina.}
\address[CAB]{Instituto Balseiro, CNEA-UNC, Av. Bustillo 9500, (8400) San Carlos de Bariloche, Argentina.}
\address[MX]{Instituto de Ciencias Nucleares, UNAM, Circuito Exterior S/N, Ciudad Universitaria, Mexico D.F. 04510, Mexico}

\thanks[cor1]{Corresponding author. E-mail: medina@tandar.cnea.gov.ar. Laboratorio Tandar, Av. Gral. Paz 1499 (1650), Buenos Aires, Argentina. Tel: +54 11 67727199. Fax: +54 11 67727131.}
\thanks[CONICET]{CONICET, Argentina.}
\thanks[CONI]{Member of Carrera del Investigador Cient\'{\i}fico, CONICET, Argentina.}

\begin{abstract}

The Pierre Auger Observatory has been designed to study the
highest-energy cosmic rays in nature ($E \geq 10^{18.5}$ eV).  The
determination of their arrival direction, energy and composition
is performed by the analysis of the atmospheric showers they
produce. The Auger Surface Array will consist of 1600 water
Cerenkov detectors placed in an equilateral triangular grid of
1.5\,km spacing. The aim of this paper is to show that the
addition of a ``small" area of surface detectors at half or less
the above mentioned spacing would allow a dramatic increase of the
physical scope of this Observatory, reaching lower energies at
which the transition from galactic to extragalactic sources is
expected.

\end{abstract}


\end{frontmatter}

\newpage

\section{Introduction and Motivations}

\begin{figure}[tbp]
\begin{center}
\includegraphics*[height=15cm,angle=-90 ]{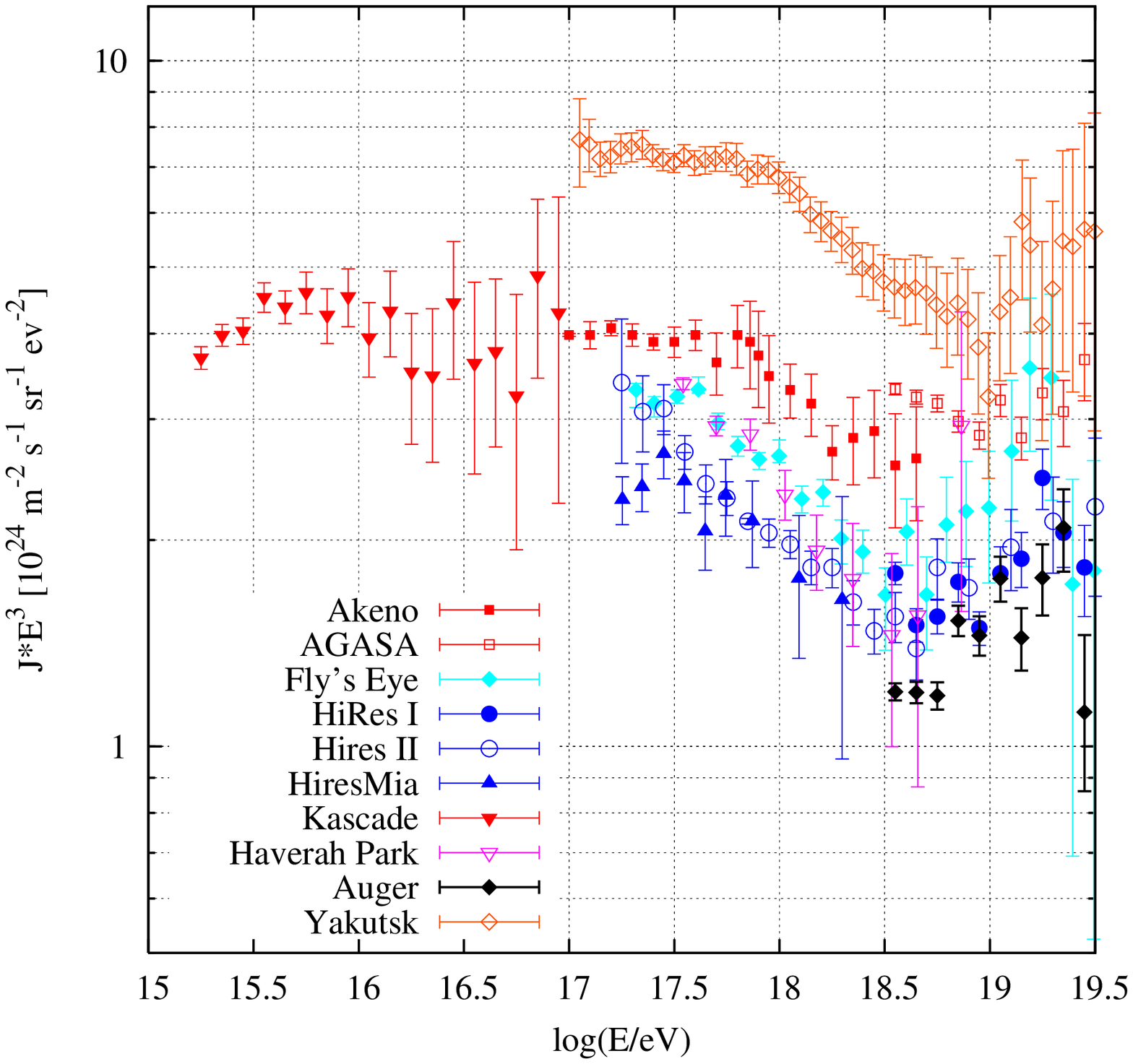}
\caption{Differential cosmic ray spectra multiplied by $E^3$,
showing the knee, the suggested second knee and the ankle features.
Compilation from KASCADE \cite{Ant}, AKENO \cite{Nag} Yakutsk
\cite{Glush}, Haverah Park \cite{Ave}, Fly's Eye Stereo
\cite{Bird03}, HiRes-MIA \cite{Abba}, HiRes I and II \cite{Abba},
AGASA \cite{Tak}, and AUGER \cite{AugerSpect}. The KASCADE data was
obtained using the QGSJET 01 interaction \cite{Kalmy} and AKENO data
is from the 1 $km^{2}$ array.} \label{fig:figSpect}
\end{center}
\end{figure}

Cosmic rays are observed in a wide range of energies spanning more
than eleven decades, from energies below 1 GeV up to more than
$10^{20}$ eV.  At lower energies the spectrum follows a simple
power law with an exponent equal to -2.7, compatible with galactic
supernova remnant acceleration of charged nuclei. The spectrum
slope becomes steeper at $\sim 3-5 \times 10^{15}$ eV, a feature
known as the \emph{knee}, where the spectral index changes from
-2.7 to -3.1. A spectrum compilation from $10^{15}$ eV (just below
the knee) to the highest detected energies is shown Fig.
\ref{fig:figSpect}. This figure clearly depicts a large systematic
uncertainty in energy calibration of $\sim 30 \%$ among the
different experiments (and even larger for the Yakutsk data).
Still, when the different experimental data are
energy-renormalized there is fair agreement in shape up to $E \leq
10^{19.5}$ eV.

The KASCADE experimental results \cite {Kamp,Agl,Ant} show evidence
that the first knee is essentially caused by a decreasing flux of
light primaries: as the atomic number increases, each
element-associated knee moves to a higher energy. The KASCADE data
in Fig. \ref{fig:figSpect} show that the experiment is finely
designed to inspect the knee region, i.e. in the $\sim 10^{15} -
10^{16.5}$ eV energy range. Although it collects data above this
energy and up to $10^{17}$ eV, it does it with larger statistical
uncertainties as it operates on the verge of its acceptance.

In the region  from $\sim 10^{17.0}$ to $10^{19.5}$ eV the spectrum
is reported to show two further traits: a break, called the
\emph{second-knee}, and a broad feature known as the \emph{ankle}.
The second-knee feature has been suggested to be a realization of
the knee for the heaviest stable elements, i.e., Fe (see for example
\cite{Haungs01}, \cite{Hoer01}). It may represent the end of the
efficiency of supernova remnant shock waves as accelerators or a
change in the diffusion regime inside our Galaxy. Fig. 1 tells that
AKENO \cite{Nag}, Yakutsk, with its latest analysis \cite{Glush},
and Fly's Eye stereo observe a clear second-knee-like feature. The
figure also shows that KASCADE, Haverah Park, AGASA, HiRes and Auger
have not enough acceptance at the second-knee region. HiRes/MIA is
at the limit of its acceptance, but there is still a hint of the
second knee in its data.

This acceptance gap will be filled by KASCADE-Grande, which will
have an exposure area ten times larger than Kascade and will attain
good statistics up to close to $10^{18}$ eV by 2009, with
composition studies up to (0.7 - 0.8)$\times 10^{18}$ eV
\cite{Haung02}. If so, it will bridge the undoubtedly very important
energy range from the first to the second knee, and hopefully cast
light on primary cosmic ray composition.

As already mentioned, the second knee might be the end of the stable
elements of the cosmic ray Galactic component dominance
\cite{Haungs01,Hoer01,Hoer02}. Therefore, at higher energies an
additional component would be required to account for the observed
flux. In Refs. \cite{Ber01,Ber02} it is conjectured that this
additional component arises from extragalactic protons, together
with some galactic heavy nuclei to conform a broad second knee
feature.

At energies beyond the second-knee feature, the spectrum clearly
exhibits a broad depression, called the \emph{ankle}. As shown in
Fig. 1, the ankle has been observed around $3 \times 10^{18}$ eV by
Fly's Eye \cite{Abu,Bird01,Bird02}, Haverah Park \cite{Ave}, HiRes
\cite{Hires}, and it is not contradicted by Auger \cite{AugerSpect}
at its preliminary stage. These results have been confirmed by
Yakutsk \cite{Glush}\cite{Ivan} and AGASA \cite{Tak} but they locate
it at a higher energy, around $10^{19}$ eV. There are at least two
physical interpretations of the ankle, intimately related to its
nature: it may be the transition between the Galactic and
extragalactic components \cite{Hillas,Wib,Allard01} or the result of
pair creation by extragalactic protons in the cosmic microwave
background \cite{Ber01,Aloi}. In the former model it is assumed that
heavy nuclei are accelerated to the ankle energy within the Galaxy,
whereas in the latter only up to the second knee and therefore the
importance is not laid on the ankle but on the broad trait which is
interpreted as a pair-production dip. It has also been suggested
that the ankle is the result of diffusive propagation of
extragalactic nuclei through cosmic magnetic fields
\cite{Kalm,Ogio}. To pinpoint the correct model both a reliable,
high quality spectrum with a well calibrated absolute energy and a
detailed composition study are required, as the pair-production dip
model relies on cosmic rays being essentially protons beyond the
second-knee \cite{Ber02,Allard01}.

In any case, the energy region spanning from $\log (E/eV) \sim 17.5$
to $\sim 19.0$ very likely comprises the transition from Galactic to
extragalactic cosmic rays.

\subsection{The Pierre Auger Observatory}
As of today the forefront experiment in the ultra-high energy cosmic
ray arena is the Pierre Auger Observatory \cite{AugerNIM}, that aims
at building two observatories, one in each hemisphere. The
construction of the Southern  Observatory started in 2000. It is
already taking data and approaching completion. It is located in the
region called ``Pampa Amarilla'', close to Malarg\"{u}e, at the south of
the Province of Mendoza in Argentina ($35.0^{\circ} - 35.3^{\circ}
S, 69.0^{\circ} - 69.3^{\circ} W$), therefore having a full view of
the Galactic center and its surroundings. Auger's two distinctive
features are its exceptional size and its hybrid nature. It spans
over an area of 3000 km$^{2}$ and is constituted by a surface array
of 1600 water $\check{C}$erenkov detectors placed on a 1.5 km
triangular grid plus 24 fluorescence detector telescopes placed in 4
buildings on the Surface Detector (SD) array periphery and
overlooking it. Consequently, it will provide a large number of
events with low systematic uncertainties.

Pertaining to this work is the Auger surface detector array. It has
a versatile trigger system \cite{Allard02} designed to operate in a
wide range of primary energies and arrival directions. Apart of the
customary Signal-over-Threshold (ST) trigger, Auger bases its SD
local trigger system on a Time-over-Threshold (ToT) trigger, which
requires the signal to be above the ToT threshold (which is much
lower than the ST threshold) during at least 325 ns in a 3 $\mu$s
interval. The array trigger requirement is satisfied when at least 3
surface detectors in a compact configuration detect a local trigger
in coincidence.  With this set of conditions the Southern
Observatory is fully efficient above $3\times10^{18}$ eV
\cite{Allard03}. For energies below this value, the detection
efficiency decreases rapidly and is composition dependent. Also, the
shower's parameters (i.e. energy, zenithal angle, and core position)
reconstruction is degraded since an average of three tanks does not
suffice to convincingly sample the shower lateral distribution
function (LDF). Detection efficiency also becomes fluctuating with
atmospheric conditions (temperature and pressure), introducing
unknown uncertainties in sky coverage.

\subsection{The Graded Infill}
In this paper it is proposed to enhance the Auger acceptance down to
$10^{17}$ eV by means of a graded infill of surface detectors
deployed at smaller distances, over an area much smaller than Auger
because of the much higher flux at lower energies. This infill
array would use the same technology of Auger surface detectors and
benefit from the existing knowledge and infrastructure at the site
(detector design, data acquisition, analysis tools, etc.). Such an extension
would allow: (a) to make a reliable measurement with low
reconstruction uncertainty of the energy spectrum from the second
knee to the ankle, (b) to make more accurate anisotropy measurements
at lower energies avoiding the many spurious effects like the
temperature dependence of the event rate and the composition
dependent efficiency, and (c) to bridge with a sizeable overlap the
data between KASCADE-Grande and Auger, thus increasing the
reliability of the results obtained.

An infill would also give an experimental handle on fluctuations at
the verge of Auger acceptance, i.e. $ E\sim 3 \times 10^{18}$ eV, as
it will allow event reconstructions with two or more non-overlapping
subsets of detectors at 1500 m distance. In a similar fashion it
would permit a very precise experimental determination of the
acceptance of the full array, which will also serve to validate
detector simulation packages. Additionally, the infill could
contribute to a better primary discrimination through a comparison
with the dependence of the acceptance on composition.

Three different infill configurations are considered, which can be
obtained by adding surface detectors to the Auger grid, which has a
spacing of 1500 m. These infills give new triangular arrays with
spacings between neighbouring detectors of, respectively, 866, 750,
and 433 m. In order to assess the impact of such an enhancement, we
study by means of simulations the dependence of the resolution for
different shower parameters (arrival direction, core position,
lateral distribution of shower particles, energy, etc.) on detector
spacing, for different primary energies, composition, and zenith
angles. We will conclude that in order to reach the above mentioned
energy range with unitary efficiency, a graded infill is required,
i.e, an array with a 750 m-grid and, in an even smaller area, an
infill of this infill, with a spacing of 433 m between SD´s.


This paper is organized as follows: section \ref{sec:Simulations}
describes the simulations of showers and detector response, as
well as the analysis methods used. Section \ref{sec:acept}
contains our acceptance calculation for different infill
configurations. In sections \ref{sec:ang_res} to \ref{sec:energia}
the improvement on reconstruction of shower parameters is
analyzed. Some considerations about the possible contribution to
composition studies are given in section \ref{sec:composition};
conclusions are presented in section \ref{sec:conclusion}.

\section{Simulations}\label{sec:Simulations}

For this work we generated a library of extensive air showers using
the Monte Carlo simulation code Aires 2.6.0 \cite{Sciut} (see also
\cite{Knapp}). Two types of primaries (proton and iron) were
considered, arriving with three characteristic zenith angles:
$0^{\circ}$, $30^{\circ}$ and $45^{\circ}$, with five different
energies: $10^{17.5}$ eV, $10^{17.75}$ eV, $10^{18}$ eV,
$10^{18.25}$ eV and $10^{18.5}$ eV. For each energy, zenith angle
and primary composition,  100 showers were simulated with a uniform
azimuthal distribution. With each of these showers, 5 events were
generated by triggering the surface array at random impact points.

\begin{figure}[tbp]
\begin{center}
\includegraphics*[height=7cm,angle=-90]{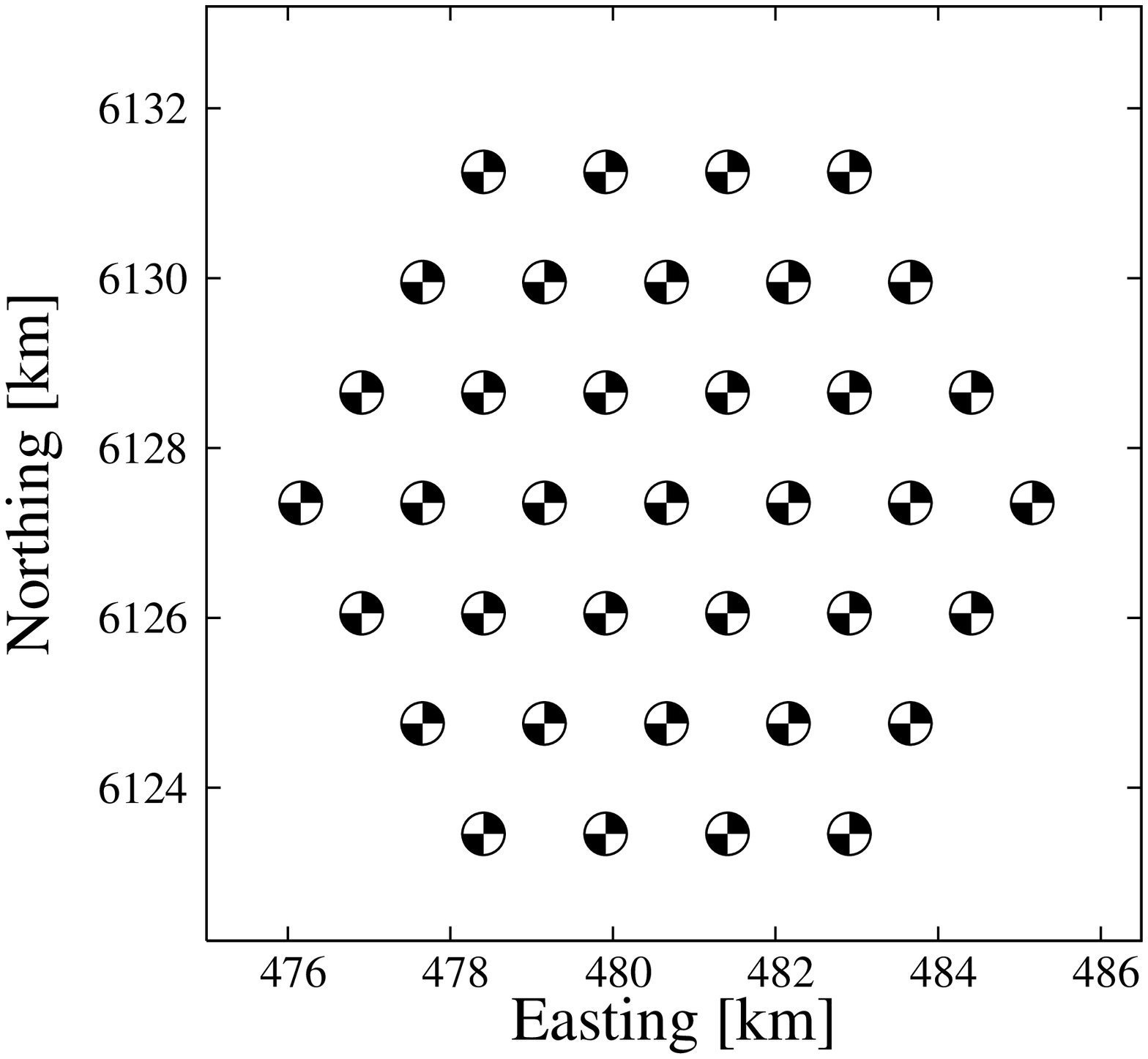}
\includegraphics*[height=7cm,angle=-90]{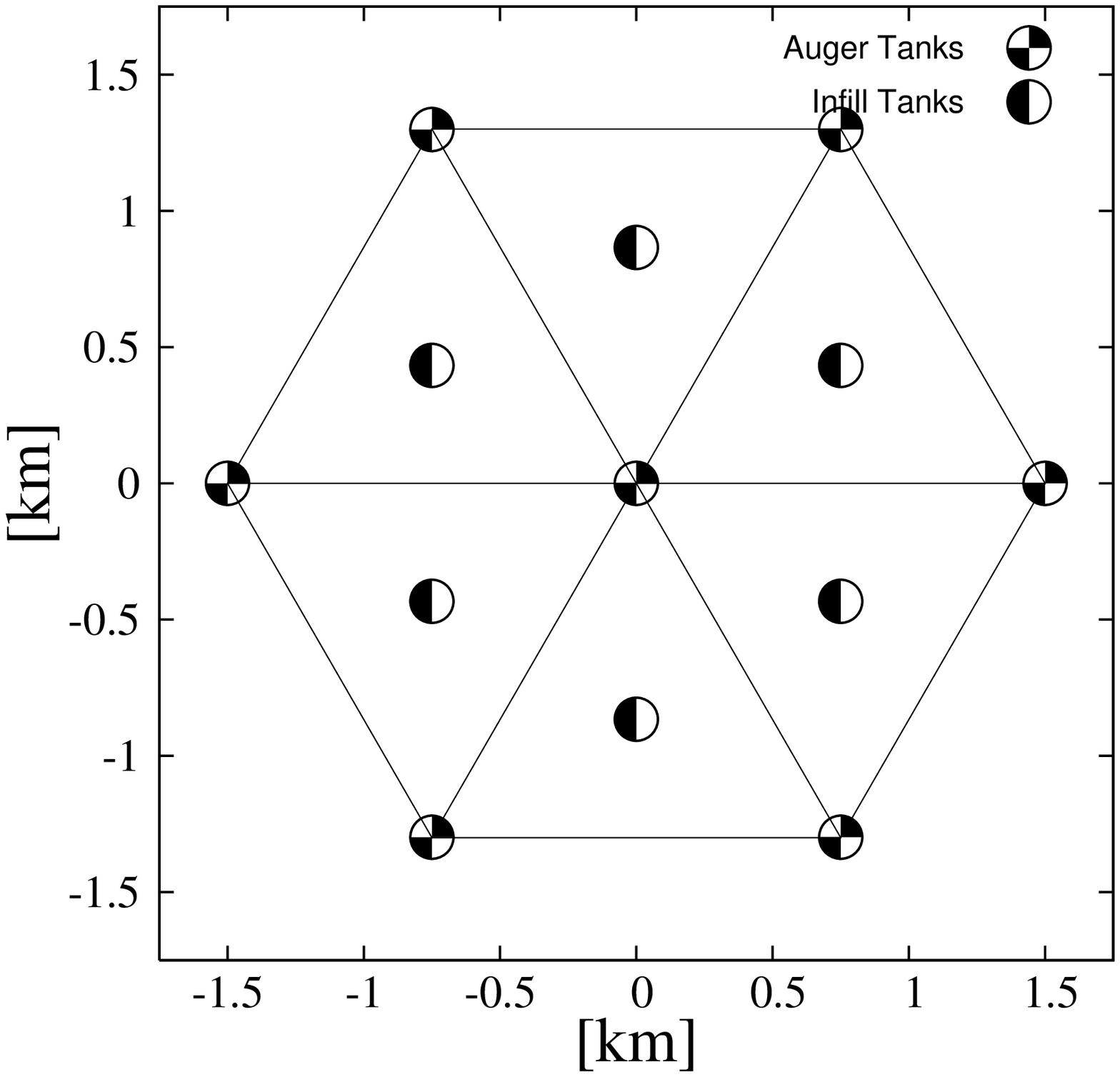}
\includegraphics*[height=7cm,angle=-90]{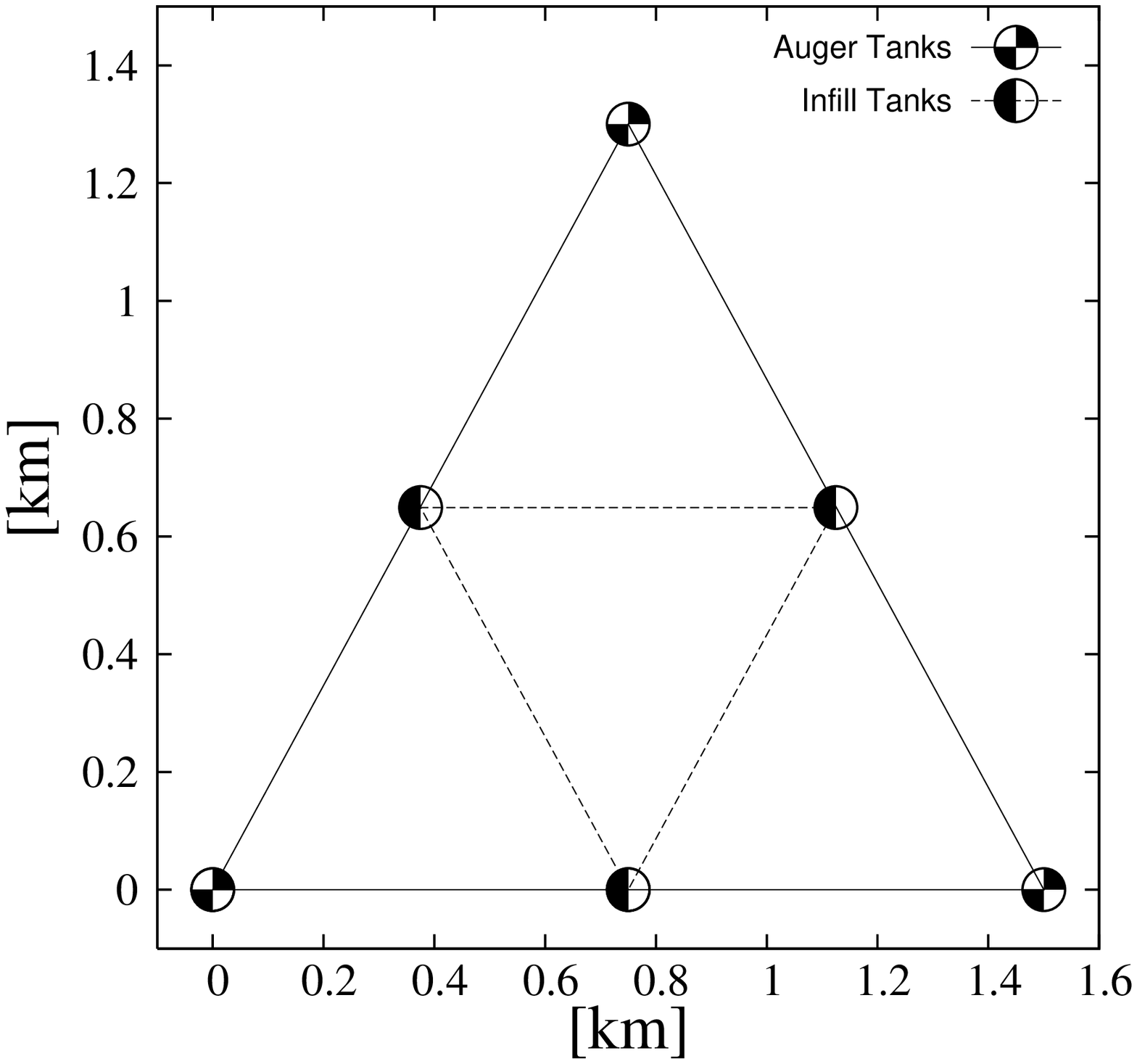}
\includegraphics*[height=7cm,angle=-90]{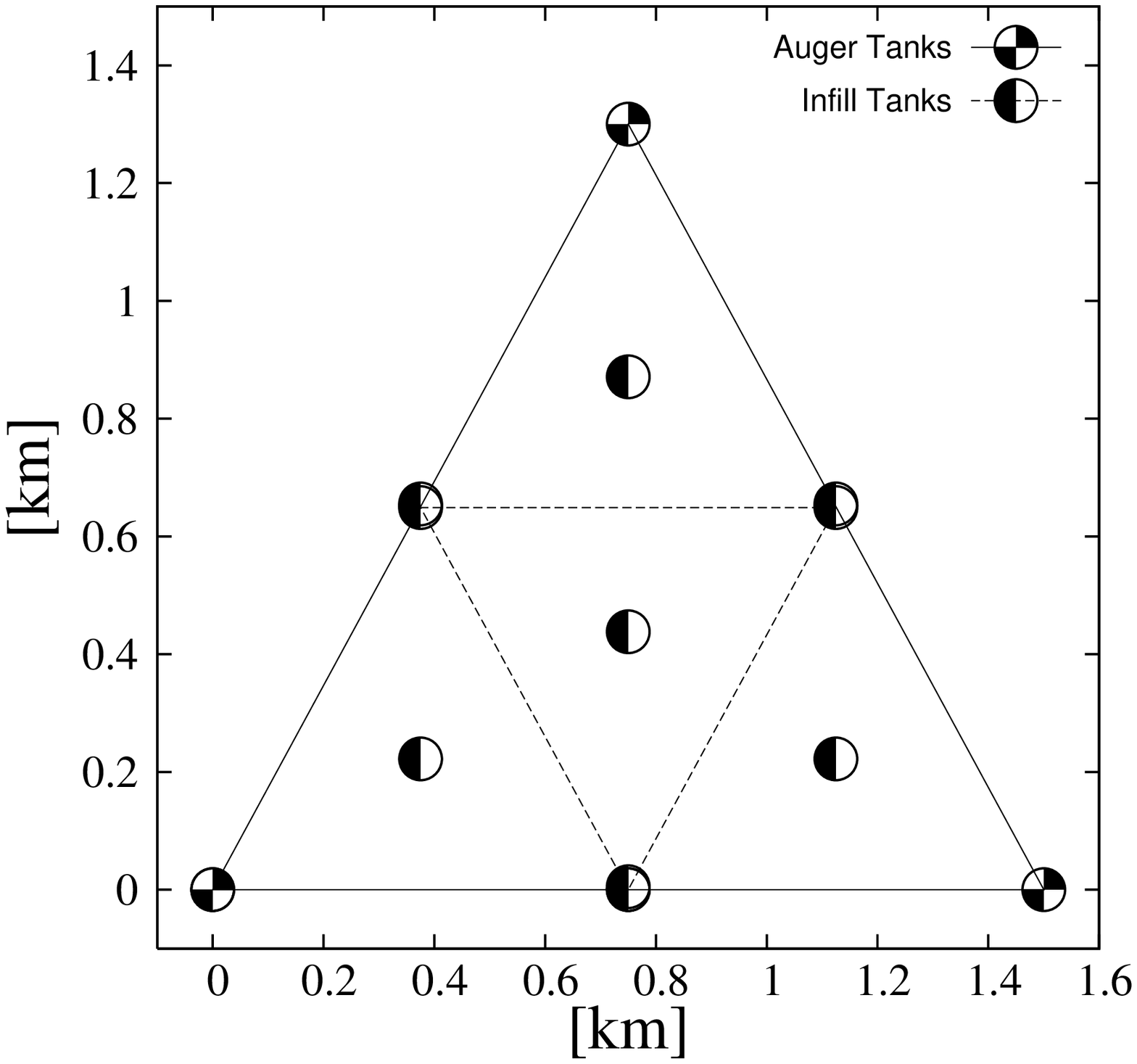}
\caption{Different possible infill configurations generated inside
the original array with 1500 m spacing (\textit{top  left}): a
detector placed at the center of each triangle (\textit{top  right})
makes a grid of 866 m. A detector added at half distance between the
original detectors (\textit{bottom  left}) results in a grid of 750
m. A detector  placed at the center of each triangle of 750 m
(\textit{bottom  right}) produces a grid of 433 m.}
\label{fig:figSim}
\end{center}
\end{figure}

Using the code SDSim (v3r0) \cite{Dag-C}, the response of an
Auger-like surface array was simulated, consisting of 37 detectors
covering a hexagonal area of 52 $km^2$ (see Fig. \ref{fig:figSim},
top left). Three different infill configurations were added to this
1.5 km triangular grid of detectors:

\begin{itemize}
\item a) One detector at the center of each triangle of 1500 m,
resulting in a triangular grid with a spacing of 866 m (Fig.
\ref{fig:figSim}, top right)
\item b) Detectors at
half the distance between Auger detectors, resulting in a triangular
grid with a spacing of 750 m  (Fig. \ref{fig:figSim},  bottom left)
\item c) Detectors at the center of each
triangle of 750 m from grid b), resulting in  a triangular grid with
a spacing of 433 m (Fig. \ref{fig:figSim}, bottom right)
\end{itemize}

All Auger trigger levels \cite{Allard02} with their relevant signal
thresholds have been used in these simulations. After all trigger
requirements are met, a full reconstruction of the arrival direction
and the lateral distribution function is performed using the event
reconstruction package Er(v3r4) provided by the Central Data
Acquisition Group of the Auger Observatory \cite{CDAS}.
Reconstructions invoke a Lateral Distribution Function (LDF)
parameterization which describes the expected integrated signal at a
distance $r$ from the shower core. We chose the functional form:

\begin{equation}\label{ec:ldf}
    LDF(r) = S_{1000} \left({r\over r_0}\right)^{-\beta+\gamma\, log(r/r_0)}
\end{equation}

where $r_0$ =1000 m and $S_{1000}$, $\beta$ and $\gamma$ are
parameters to be adjusted.

At the lowest energy considered ($3\times10^{17}$ eV) the number of
showers that are able to trigger the Auger 1500 m-array is marginal
and, correspondingly, the determination of the resolution of shower
parameters is physically meaningless.

\begin{figure}[tbp]
\begin{center}
\includegraphics*[height=10cm,angle=-90]{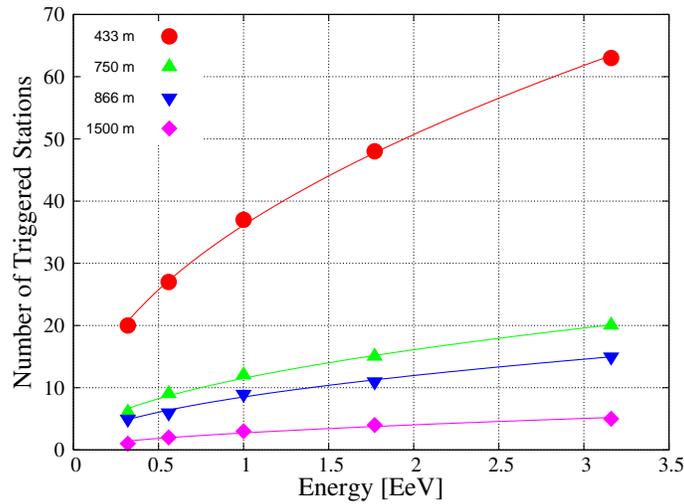}
\caption{Average number of triggered stations per event for
different spacings and energies, for iron primary with a zenithal
angle of $\theta = 30^{\circ}$.} \label{fig:figNdetE}
\end{center}
\end{figure}

The array acceptance and the arrival direction and LDF
reconstruction accuracy depend strongly on the number of stations
with signal above threshold. Fig. \ref{fig:figNdetE} shows the
number of triggered stations versus primary energy for different
infill spacings. It is seen that, as the shower area on the ground
increases with increasing energy, the number of triggered detectors
gets correspondingly larger. For a given detector spacing, this
number is expected to be proportional to $E^{2/\beta}$, where
$\beta$ is the slope of the LDF. The values and fits shown in Fig.
\ref{fig:figNdetE} correspond to iron primaries.

\section{Acceptance Determination}\label{sec:acept}

The shower detection efficiency of a surface array depends crucially
on the spacing between detectors and on the trigger configuration,
i.e., the number of detectors that are required to have signal above
a certain threshold for the event to be recorded. The surface
detector array of the Pierre Auger Observatory, with a spacing of
1500 meters between neighbouring detectors, is nearly fully
efficient for showers of more than 3 EeV, if a 3-detector trigger
configuration is required \cite{Allard02}. This value has been
obtained by Monte Carlo simulations of the detector response.

The instantaneous acceptance of the surface array is given by:
\begin{equation}\label{ec:accept}
    Acc(E)_{N+}=\int P_{N+}(\vec{r},E,\theta)\cdot \cos(\theta)\cdot d\vec{r}^{2}
            \cdot 2\pi \cdot  \sin(\theta) \cdot d\theta
\end{equation}

where $P_{N+}$ is the probability of having a positive trigger with
at least $N$ detectors for a shower of energy $E$ hitting the ground
in the position $\vec{r}$, with zenithal arrival direction $\theta$.
In the full acceptance regime, i.e. when $P_{N+}=1$, the acceptance
of an array of area $A$, and for a maximum zenith angle of
$\theta_{max}$, is  $Acc_{1}=A \pi \sin^{2}(\theta_{max})$. For what
follows, $\theta_{max}$ is set to 60$^{\circ}$.





To calculate the probability $P_{N+}$, we need to consider the local
trigger probability for a single detector (say, the \textit{i}-th
detector of the array). To a good approximation, this probability is
determined solely by the total expected integrated signal $S_i$ of
the station. Thus, if the $i$-th local station is located at $\vec
r_i$ and has an expected signal $S_i$, its trigger probability is
given by

\begin{equation}\label{ec:pdsdef}
    P(S_{i})=P(LDF(|\vec{r}_{i}-{\vec r}|,E,\theta)).
\end{equation}

Here we are making the general assumption that the expected signal
at a certain distance from the shower core $\vec r$ can be modelled
by a Lateral Distribution Function (LDF) (see section \ref{ec:ldf}).

To obtain $P(S)$, a set of showers in the 10$^{16}$ - 10$^{19}$ eV
energy range with zenith angles up to 60$^{\circ}$ were simulated,
together with the corresponding detector response to obtain the
signals in the detectors. These simulated signals, which include the
fluctuations both in the shower development as well as in the
detector response, are used to reconstruct the shower core position,
energy and incidence angle by adjusting an LDF to the signals in the
triggered detectors.

From a large number of simulated showers with different core
positions the $P(S)$ can be inferred as

\begin{equation}\label{defpdes}
P(S) = \frac{N_{T+}(S)}{N_{ON}(S)},
\end{equation}

where $N_{T+}(S)$ is the number of triggered stations with an
expected signal $S$, and $N_{ON}(S)$ is the total number of active
stations in that signal bin.  Given this $P(S)$, the integral
(\ref{ec:accept}) can be calculated numerically.

\begin{figure}[tbp]
\begin{center}
\includegraphics*[height=10cm,angle=-90]{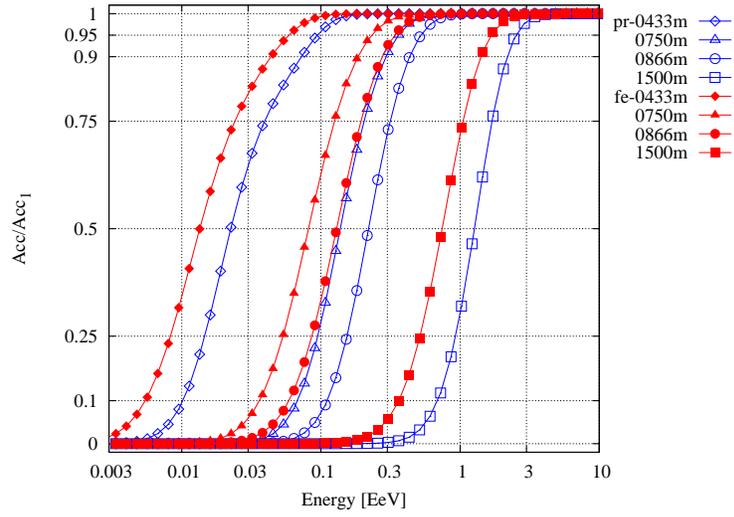}
\end{center}
\caption{Relative acceptance for different infill configurations for
proton (\textit{open symbols}) and iron (\textit{closed symbols}). }
\label{fig:figAccept}
\end{figure}

Fig. \ref{fig:figAccept} shows the resulting relative acceptance
$Acc/Acc_1$ as a function of energy, for protons (\textit{open
symbols}) and iron (\textit{closed symbols}) and for the different
detector spacings under consideration. It can be observed that a
separation of 750 m between detectors guarantees a detection
efficiency of 95$\%$ for proton (iron) showers with an energy of
$3.6\times10^{17}$ eV ($1.8\times10^{17}$ eV). With a detector
spacing of 433 m, a similar acceptance can be obtained for showers
of as low as $9\times10^{16}$ eV and $4\times10^{16}$ eV for proton
and iron primaries, respectively. According to this calculation, at
3 EeV the Pierre Auger array would attain full acceptance for a
3-detector trigger, irrespective of composition, in agreement with
ref. \cite{Allard03}.

Considering a differential flux of cosmic rays ${d\phi \over dE}$
following a power law with spectral index $-2.84$ as quoted by Auger
in \cite{AugerSpect}:
\begin{equation}
E {d\phi \over dE} \, =\, 30.9  \left(E\over EeV\right)^{-1.84}\,
km^{-2} \cdot yr^{-1} \cdot sr^{-1}
\end{equation}

the number  of expected events in one year, for events with energy
larger than $E_0$ with a zenith angle below $\theta_{max} =
60^{\circ}$, for a detector of area $A$ is given by:
\begin{equation}
N = 795 \cdot \left[ {A\over 20 km^2} \right] \left[ {t\over yr
}\right] \left[ {E_0\over 1 EeV}\right]^{-1.84}
\end {equation}

Consequently, an infill with an effective area of $\sim$20 $km^2$
(for well-contained events), which would comprise a total of 64
surface detectors separated by 750 m, could accumulate a
statistically significant number of events with a relatively small
effort: only 44 detectors beyond the regular Auger array would be
required, which amounts to less than 3$\%$ of the full Auger
Observatory. These detectors would operate in an energy region in
which Auger ceases to be fully efficient (which renders the
reconstruction not reliable), down to energies in which the regular
Auger array does not trigger any more.

As already mentioned, to detect showers with full efficiency at
energies as low as $10^{17}$ eV, a denser infill, with a detector
spacing of 433 m, is required. A statistically significant number of
events can be recorded with the addition of only 10 detectors with
such a spacing, covering an effective area of 1.6 $km^2$.

\section{Angular Resolution}\label{sec:ang_res}

The accuracy in the determination of the arrival directions of
cosmic rays is fundamental in the search for their origin and in the
study of anisotropies, although it is noted that only neutral
particles are not deflected during their propagation in magnetic
fields. We define the angular reconstruction uncertainty as the
space angle $\Theta$ subtended by the real ($\hat{R}_{real}$) and
reconstructed ($\hat{R}_{rec}$) directions, being $cos(\Theta)
=\hat{R}_{real}\cdot\hat{R}_{rec}$.

\begin{figure}[tbp]
\begin{center}
\includegraphics*[height=8cm,angle=-90]{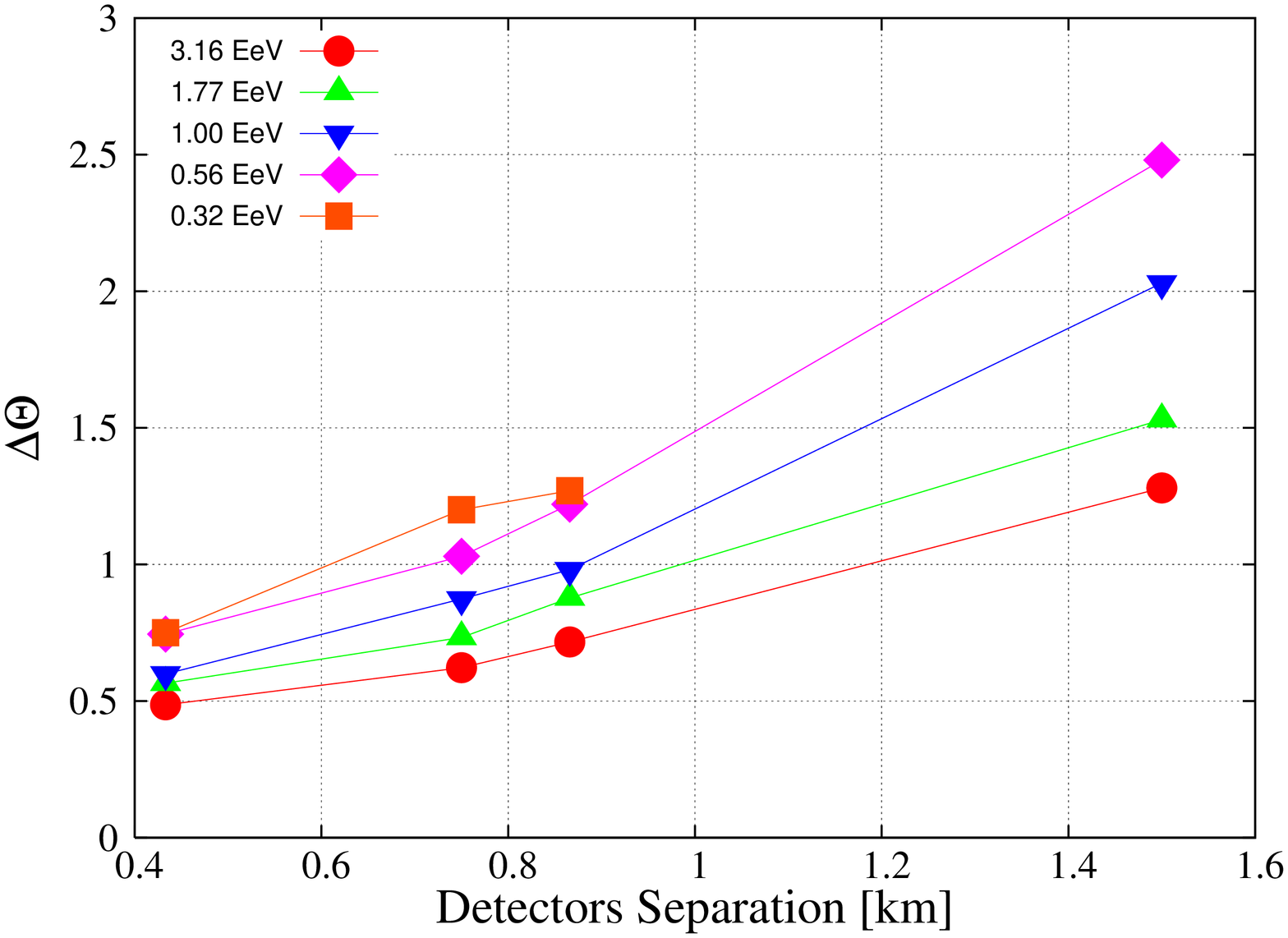}
\includegraphics*[height=8cm,angle=-90]{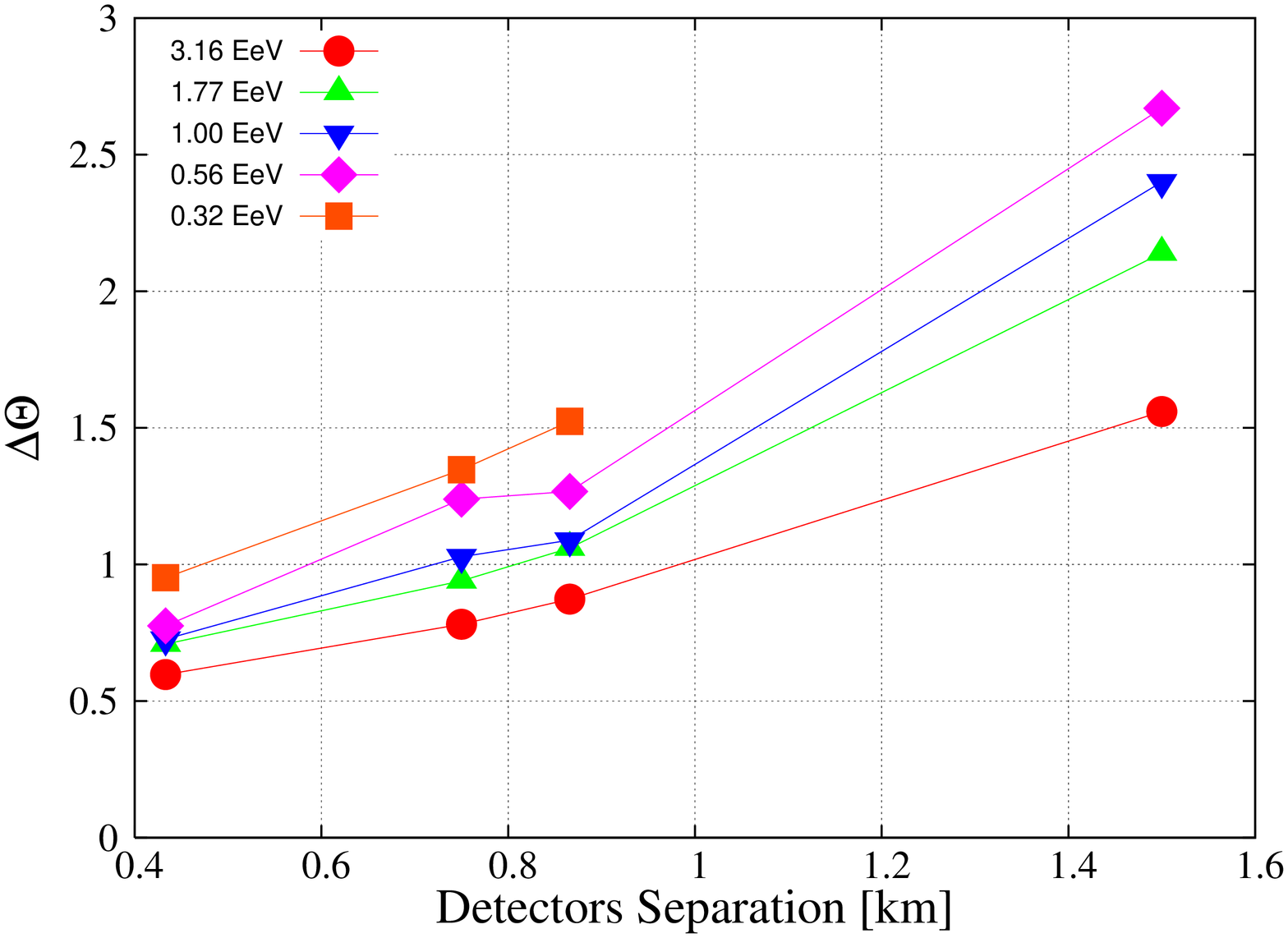}
\caption{Space angle uncertainty as a function of detector spacing
for iron (\textit{left}) and proton (\textit{right}) for $\theta =
30^{\circ}$. Lines are drawn only to guide the eye.}
\label{fig:figAngle}
\end{center}
\end{figure}

In Fig. \ref{fig:figAngle} we present the 68$\%$ confidence level
for the arrival direction reconstruction uncertainty,
$\Delta\Theta$, as a function of detector spacing and shower energy.
Results are presented for iron (\textit{left}) and proton
(\textit{right}) at different injected energies. It can be seen that
the angular resolution for events detected by a 750 m infill array
will be 2 times better than for a 1500 m-array, giving values close
to $1^{\circ}$ for proton and iron at $10^{18}$ eV.

\begin{figure}[tbp]
\begin{center}
\includegraphics*[height=10cm,angle=-90]{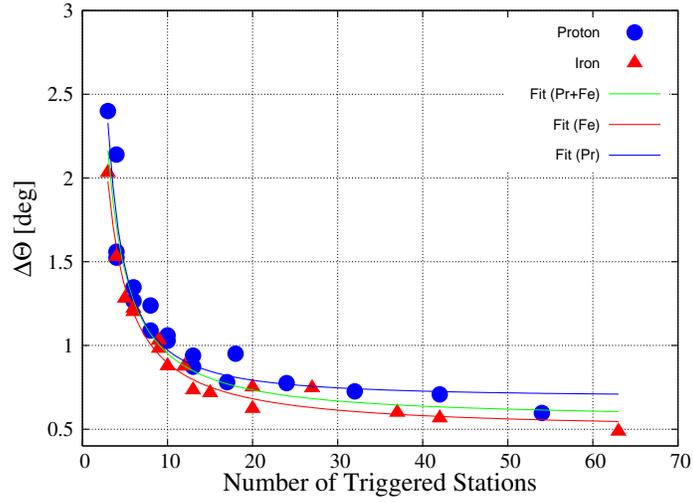}
\caption{Space angle uncertainty as function of the mean number of
triggered stations for both primaries: proton (\textit{circles}) and
iron (\textit{triangles}) for all considered energies and detector
spacings. These results correspond to $\theta = 30^{\circ}$.}
\label{fig:figNdetAng}
\end{center}
\end{figure}

It is worth analyzing the dependence of $\Delta\Theta$ on the number
of triggered detectors $N_{det}$, as shown in Fig.
\ref{fig:figNdetAng} for proton and iron primaries and for all
considered values of primary energy and detector spacing. As
expected, as the average number of triggered stations participating
in the reconstruction increases, the resolution of the arrival
direction improves. This behavior is seen for both primaries
considered. The full lines represent a fit to the data for proton
(\textit{blue}), iron (\textit{red}) and for both (\textit{green}).
The best fit for the complete set of data is given by $\Delta\Theta
= 5.8^{\circ} \times N_{det}^{-1.2} + 0.6^{\circ}$.

As the trigger conditions used in our simulations are the same as
for the Auger surface detector \cite{Allard02}, the dependence of
the angular resolution on the number of triggered stations for the
1500 m-array may be compared to that obtained by the Auger SD
\cite{ANGRESAUGER}: the values of $\Delta\Theta$ corresponding to 3,
4 and 5 stations ($2^{\circ}$, $1.6^{\circ}$ and $1.2^{\circ}$,
respectively) are in good agreement with the results presented in
\cite{ANGRESAUGER}, where it is found that for the 3-fold events,
Auger has an angular resolution of about $2^{\circ}$, for 4-fold
events of $1.7^{\circ}$ and for 5-fold events of $0.9^{\circ}$.

It is worth comparing the obtained angular resolution with that
quoted by other experiments designed to operate in a similar energy
range. The 750 m-infill is expected to give an arrival direction
resolution of $1^{\circ}$ at $10^{18}$ eV while AKENO achieves a
pointing accuracy of $3^{\circ}$ \cite{Haya}. At $10^{17}$ eV,
Kascade-Grande \cite{Haungs01} is expected to have an angular
resolution of $0.3^{\circ}$ and IceTop \cite{Stanev} claims an
accuracy of $0.6^{\circ}$. These values should be compared to the
proposed 433 m-infill which will operate in this energy region and
will have an angular resolution smaller than $0.7^{\circ}$.

\section{Core Position Resolution}\label{sec:core}

A very important parameter for the reconstruction of shower geometry
and energy is the core position, i.e. the intersection point of the
shower axis with the ground. A first estimation is given by the
barycenter of the 3 highest-signal detectors. This value is used as
input for an iterative fitting process in which the lateral
distribution of the detector's signal is adjusted.

To estimate the expected uncertainty on core reconstruction for
different configurations, we evaluate the difference between the
real and reconstructed core positions:

\begin{equation}
    \Delta Core = \sqrt{(x_{rec}-x_{real})^2+(y_{rec}-y_{real})^2}
    \label{Ec:deltacore}
\end{equation}

where $x_{rec}$, $y_{rec}$ are the coordinates of the shower core
(on the array plane) obtained from the full reconstruction process,
and $x_{real}$, $y_{real}$ are the input values for each of the
simulated showers. We define the core position resolution (for each
energy, angle and primary type considered) as the value at which the
integral of the $\Delta Core$ distribution is 68\% of the total
integral.

\begin{figure}[tbp]
\begin{center}
\includegraphics*[height=8cm,angle=-90]{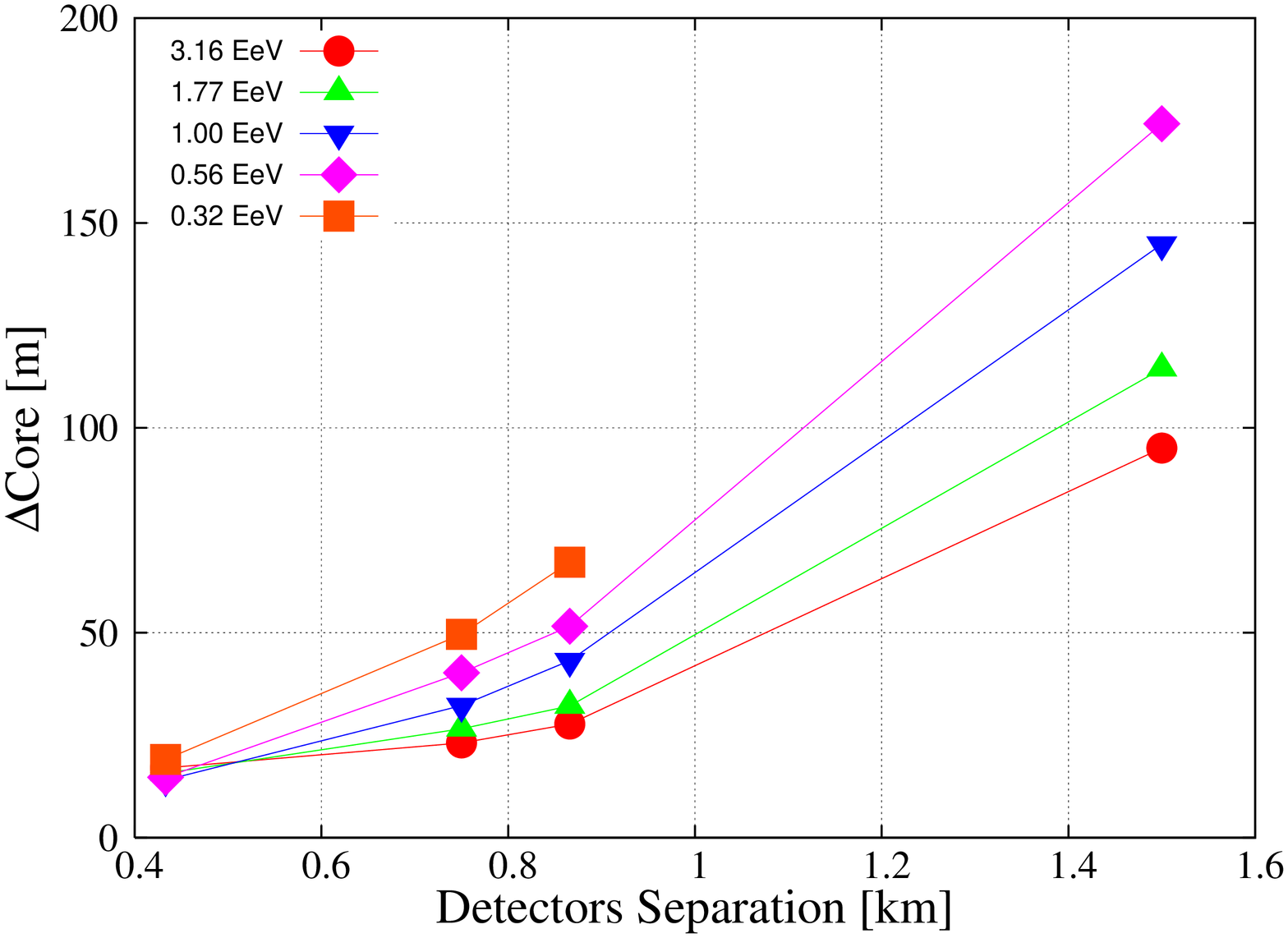}
\includegraphics*[height=8cm,angle=-90]{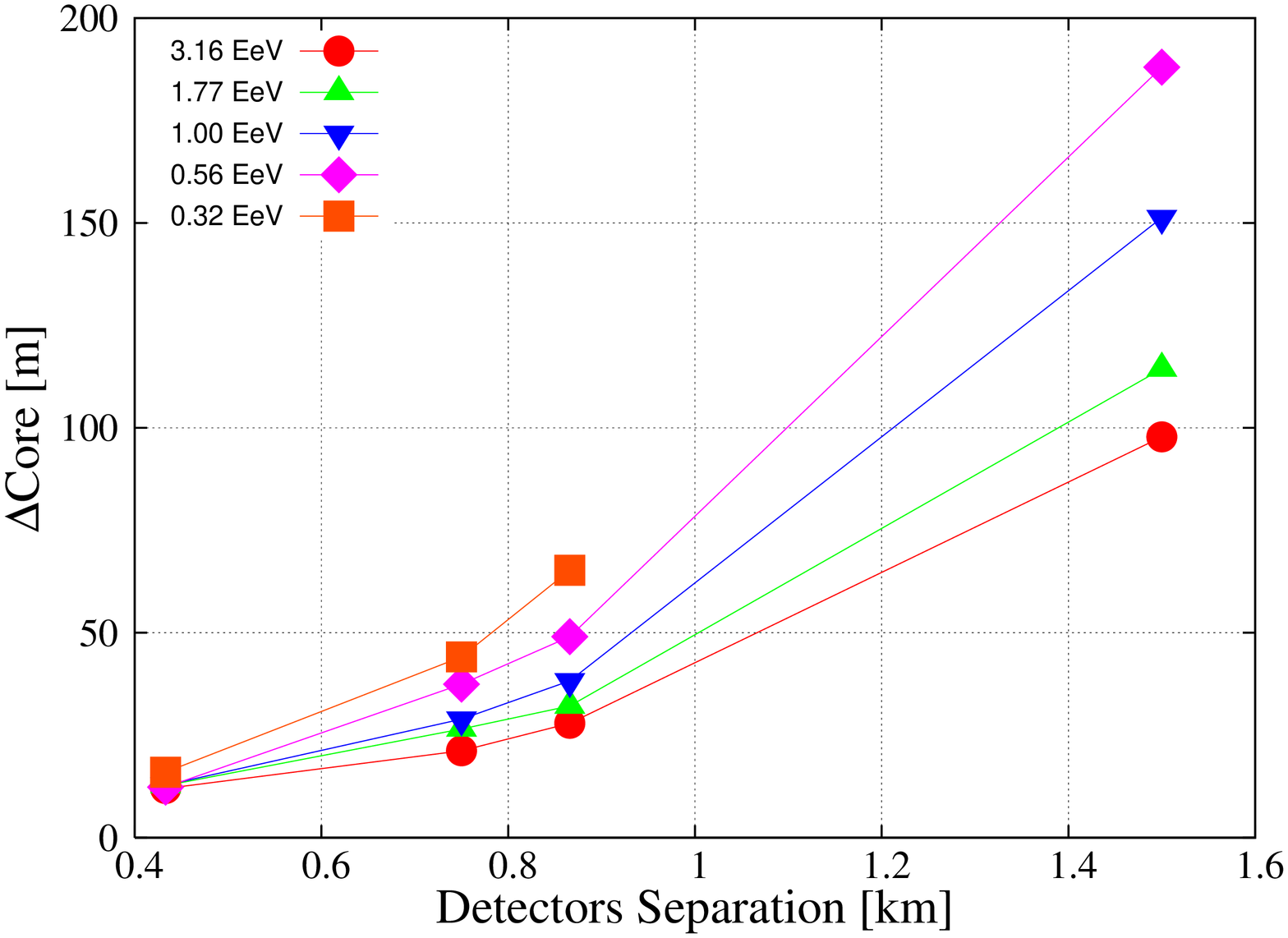}
\caption{Core position resolution as a function of detector spacing
for both primaries: iron (\textit{left}) and proton (\textit{right})
for $\theta = 30^{\circ}$. Lines are drawn only to guide the eye.}
\label{fig:figCore}
\end{center}
\end{figure}

The main results of this section are shown in Fig.
\ref{fig:figCore}. Since the lateral distribution function decreases
rapidly with distance, the presence of stations near the core plays
a fundamental role in its position determination. This situation is
favored with smaller spacings between detectors: as the distance
between detectors decreases, the core position is better determined.
Note that for the Auger spacing of 1500 m the empirical RMS value
found in \cite{S1000} is $\sim$102 m, which is in good agreement
with our results. Shower development for proton primaries present
more fluctuations than iron and this is reflected producing bigger
errors. Almost the same behavior as depicted in Fig.
\ref{fig:figCore} for showers with $\theta = 30^{\circ}$ was
observed for vertical and more inclined ($45^{\circ}$) showers.

At 433 m spacing, the number of triggered detectors is large enough
($\sim$ 20, see Fig. \ref{fig:figNdetE}) to have a core resolution
of $\sim$ 15 m irrespective of the considered energy and primary
type. Regarding the 750 m infill, it is seen that the core position
resolution improves by a factor of $\sim$ 4 with respect to Auger.
This will have an important impact on the reconstruction, giving a
better angular and energy resolution, as the LDF fitting is also
more accurate.

\begin{figure}[tbp]
\begin{center}
\includegraphics*[height=10cm,angle=-90]{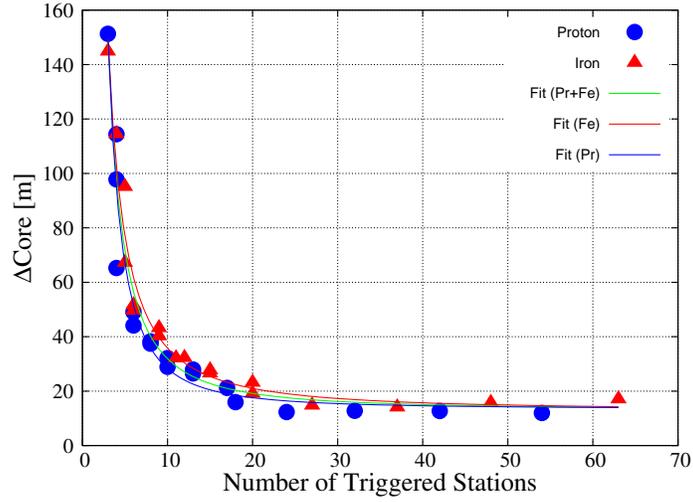}
\caption{Core position resolution as a function of the mean number
of triggered stations for both primaries: proton(\textit{circles})
and iron (\textit{triangles}), for all considered energies and detector
spacings and $\theta = 30^{\circ}$.}
\label{fig:figNdetCore}
\end{center}
\end{figure}

Fig. \ref{fig:figNdetCore} shows the core position resolution as
function of the average number of triggered detectors, for both
primary types, and all considered energies and spacings, at $\theta
=30^{\circ}$. The lines represent the best fit to the values
obtained for the different primaries (\textit{blue} for proton and
\textit{red} for iron) and for both (\textit{green}). The core
position dependence with the average number of triggered stations
can be adjusted by $\Delta Core =$ 820 m  $\times N_{det}^{-1.6} +
13 $ m.

\section{S(600) Determination}\label{sec:energia}

For the events registered by a ground array of surface detectors,
the determination of the energy consists of two steps. First, an
energy dependent parameter called $S(r_{0})$ is assigned to a given
shower. This parameter is the time-integrated signal expected in a
surface detector placed at a distance $r_{0}$ from the shower axis.
Its value is obtained by fitting an empirical LDF, with a predefined
functional form, to the observed lateral signal distribution and
interpolating its value at the distance $r_{0}$ from the core. The
second step involves an energy calibration of $S(r_{0})$. In a
hybrid detector such as Auger this can be performed either by
resorting to hybrid events, for which both the fluorescence and
surface data can be well reconstructed independently \cite{CDAS}, or
by obtaining the conversion function from Monte Carlo simulations.

The choice of the parameter $r_{0}$ for the energy conversion is
directly related to detector spacing and primary energy considered.
It has been proven \cite{S1000} that $r_0 =$ 1000 m is the optimal
value for a grid with 1.5 km spacing and for energies larger than
$\sim 5 \times 10^{18}$ eV. However, since our goal is to reach
lower energies with a smaller detector spacing, the chosen parameter
is instead $S(600)$, i.e.,  the time-integrated signal expected in a
detector at 600 meters from the shower core.

The accuracy in the determination of the parameter $S(r_{0})$
depends on the detector resolution and on sampling fluctuations.
Additionally, this accuracy has also a strong dependence on shower
to shower fluctuations which are intrinsic to the shower development
and can not be eliminated, as well as on shower direction and
relative core position \cite{S1000}. Moreover, the non-linear
interpolation used in the determination of $S(r_{0})$ increases even
further its uncertainty. It was found empirically \cite{LDFMEAS}
that for the Auger grid the statistical uncertainty on $S(1000)$
reaches 10$\%$ (corresponding to an error of 50 m for the core
location), the same order of magnitude as the sampling contribution.
These values were obtained for showers with E $\geq 5\times 10^{18}$
eV.

\begin{figure}[tbp]
\begin{center}
\includegraphics*[height=8cm,angle=-90]{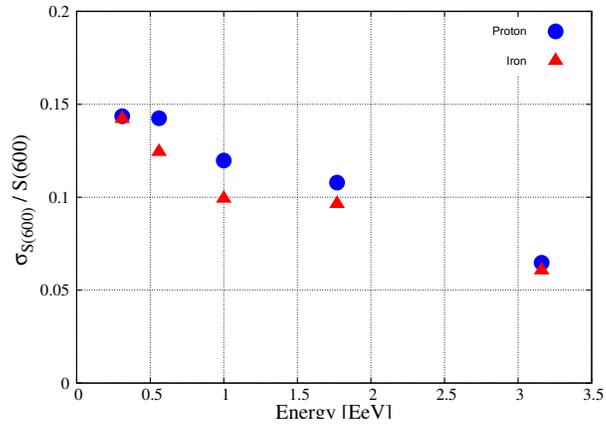}
\caption{Relative dispersion on S(600) defined as
$\sigma_{S(600)}/S(600)$ for iron (\textit{triangles}) and proton
(\textit{circles}) as function of input energy for 750 m spacing and
for $\theta = 30^{\circ}$  } \label{fig:figS600}
\end{center}
\end{figure}

In Fig. \ref{fig:figS600} we plot the ratio
$\sigma_{S(600)}/S(600)$, where $\sigma_{S(600)}$ is the dispersion
of the reconstructed S(600) distribution, for both primaries: iron
(\textit{triangles}) and proton (\textit{circles}), for different
injected primary energies. When the distance between detectors is
reduced by a factor of two, the accuracy of the new energy parameter
S(600) for $E = 10^{18}$ eV is $\sim 10 \%$ for iron primaries
whereas for protons this value is about 13 $\%$.

\section{Primary Composition}\label{sec:composition}

In the same way as the cosmic ray's energy and arrival direction are
better determined using the information provided by an infill array,
it is expected that other variables carrying information about the
identity of the primary particles would behave similarly. For
example, for surface detectors, it is known that the number of muons
at a certain distance of the shower core, the slope of the LDF, the
arrival time profile of shower particles, and the radius of
curvature of the shower front can be employed as indicators of the
primary composition \cite{Supa}. The infill allows not only a more
precise shower reconstruction as shown in the previous sections, but
also helps towards composition analyses down by another decade in
energy, completely covering the second knee and the ankle. As an
example, we studied the distribution of the LDF slope and the shower
front curvature for different characteristic energies and both
primary types, with shower and detector simulations for $\theta =
30^{\circ}$.

An estimator of the power to discriminate different primaries is
given by the parameter $\eta$ defined as:
\begin{equation}
    \eta=\frac{|Med(p)-Med(Fe)|}{\sqrt{\sigma_{p}^{2}+\sigma_{Fe}^{2}}}
    \label{eq:eta}
\end{equation}
where $Med(p)$ and $Med(Fe)$ are the medians of the distributions
for proton and iron respectively, $\sigma_{p/Fe}$are the dispersions
of the distributions \cite{Supa}.

\begin{figure}[tbp]
\begin{center}
\includegraphics*[height=8cm,angle=-90]{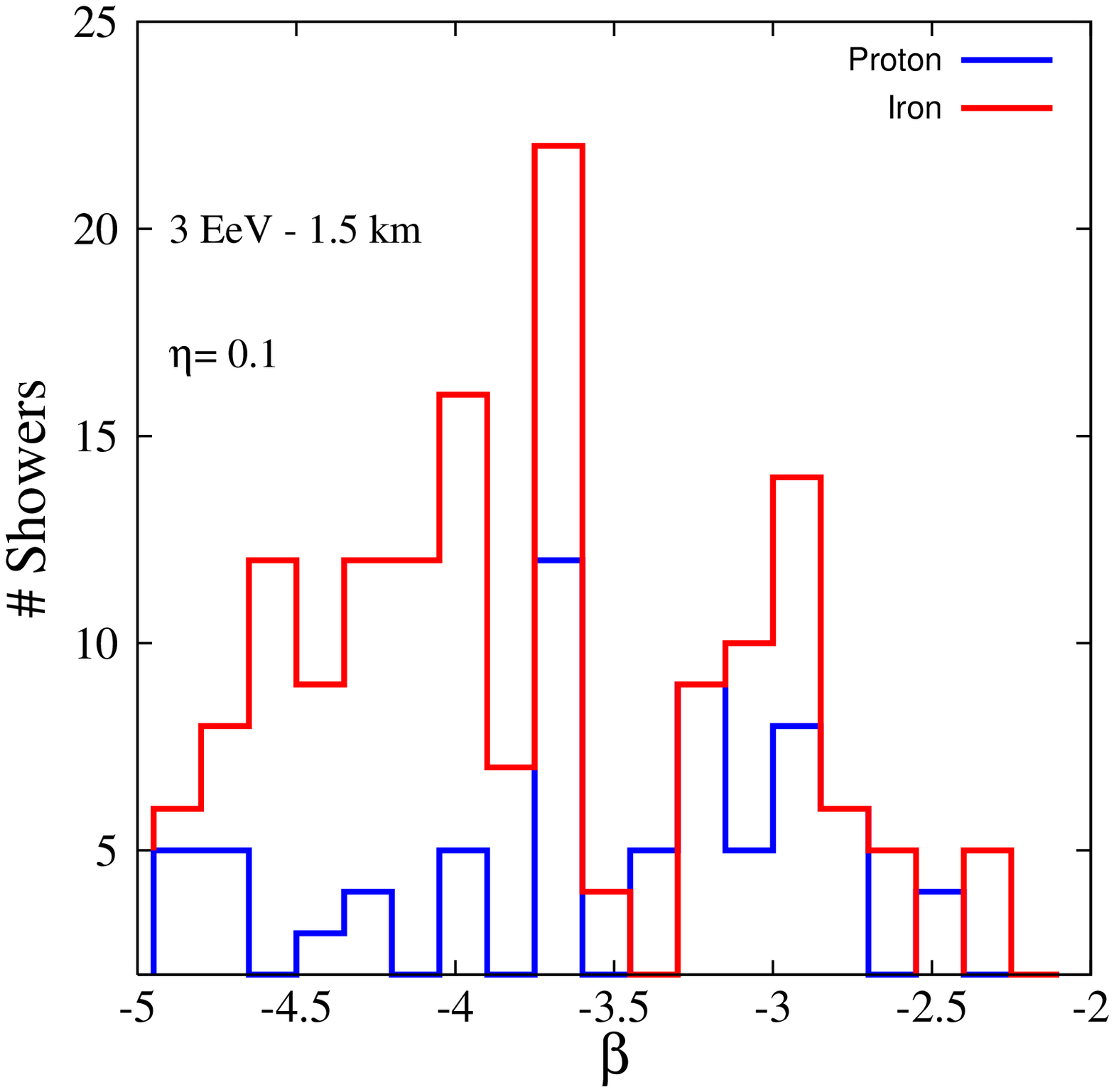}
\includegraphics*[height=8cm,angle=-90]{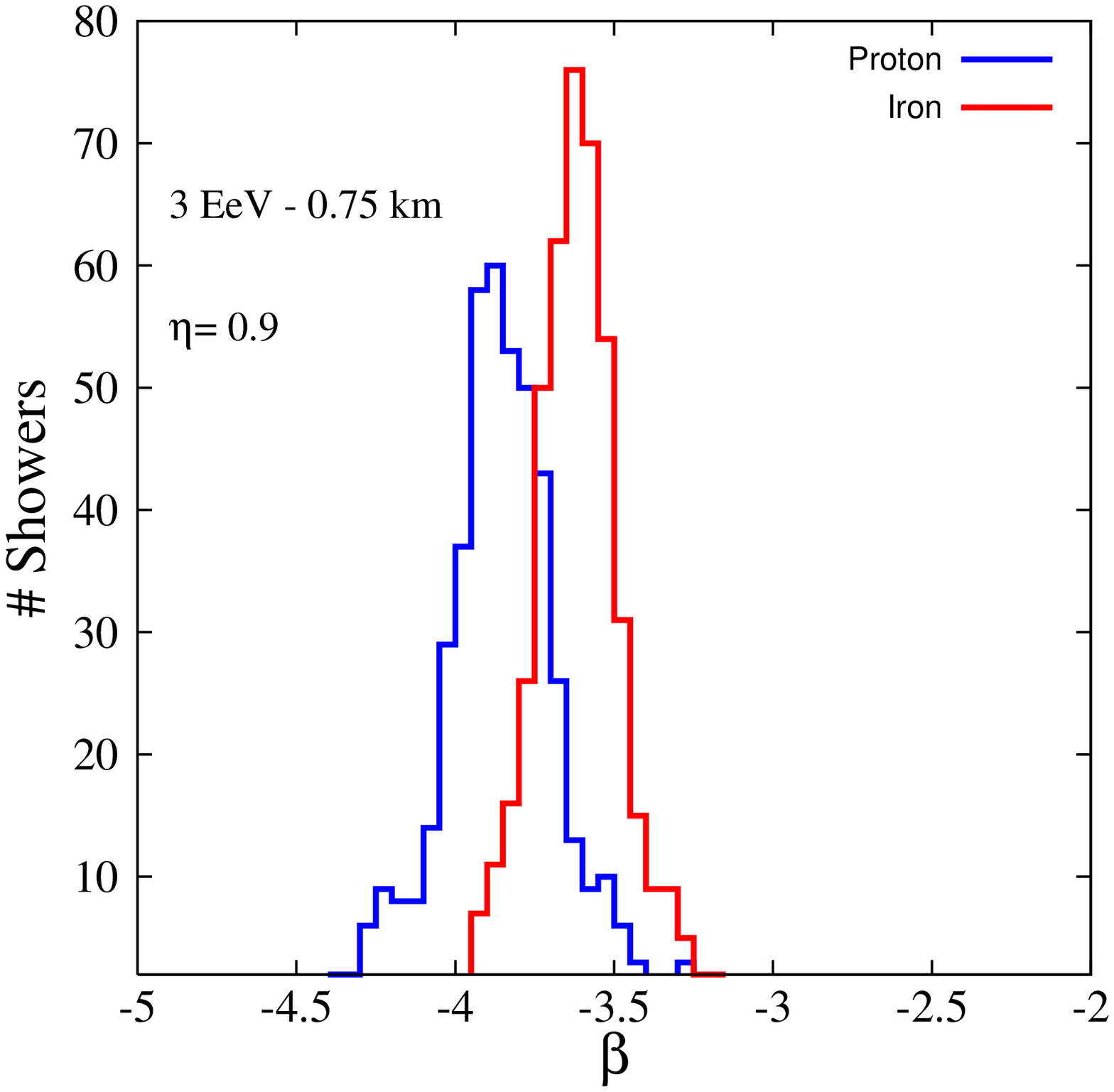}
\includegraphics*[height=8cm,angle=-90]{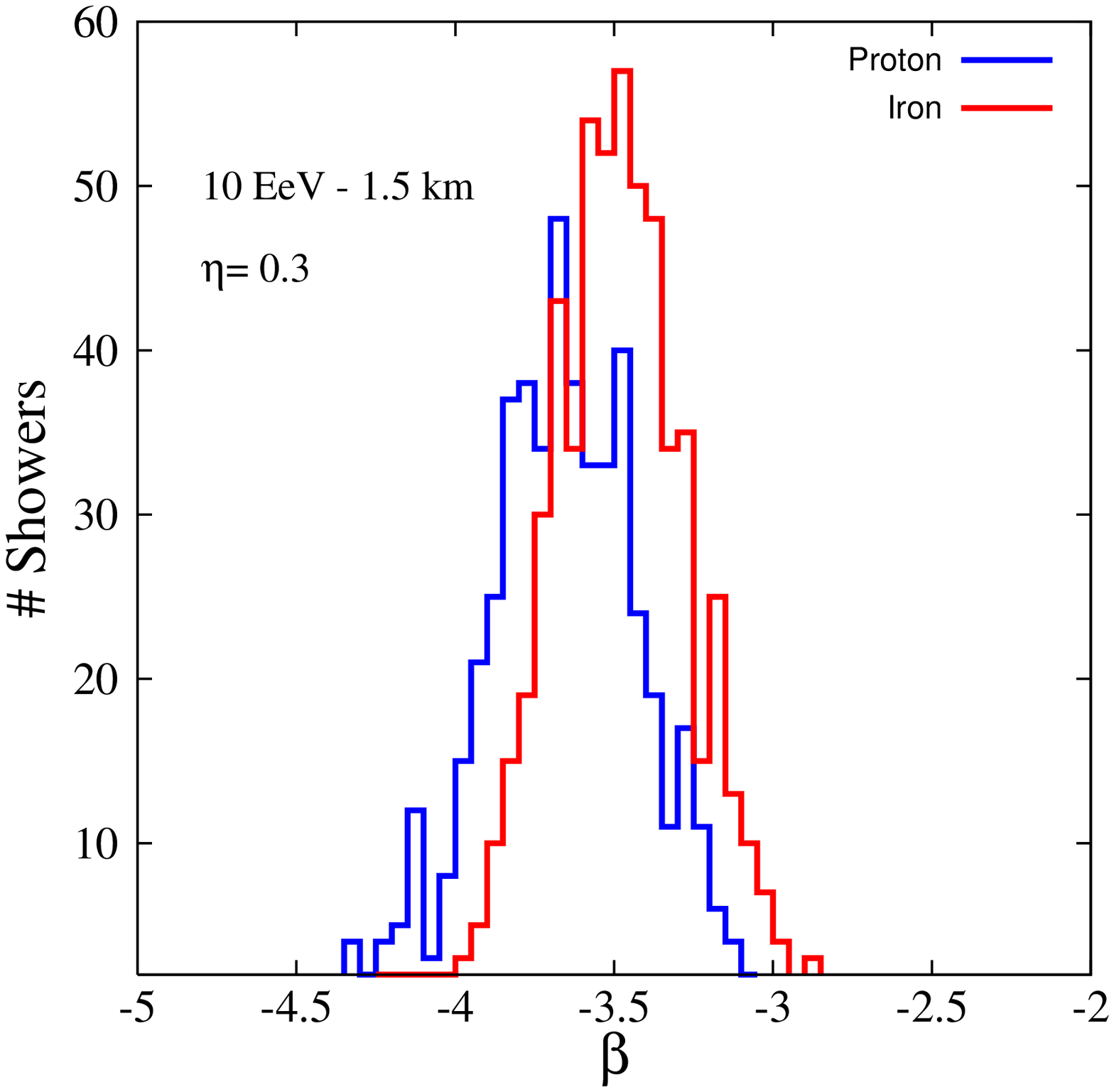}
\includegraphics*[height=8cm,angle=-90]{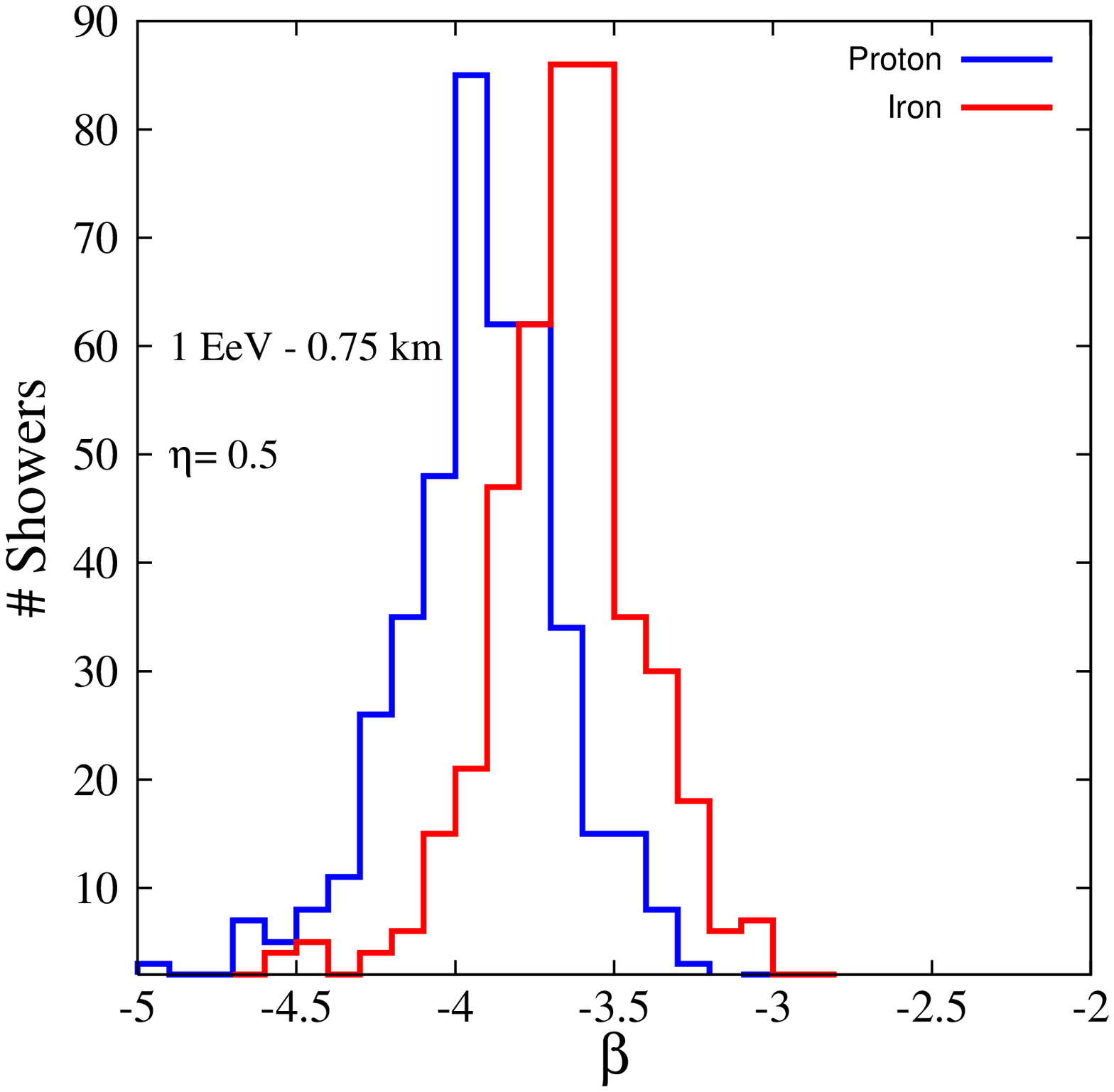}
\caption{\textbf{Top}: Distributions of $\beta$ (slope of the LDF)
for iron and proton at the limit of Auger acceptance,
i.e. $3\times10^{18}$ eV, for a spacing of 1500 m (\textit{left}) and
an infill of 750 m (\textit{right}). \textbf{Bottom}: Distributions
of $\beta$ for both primaries at $10^{19}$ eV for a Auger 1500 m-array
(\textit{left}) and at $10^{18}$ eV for a detector spacing of 750 m
(\textit{right}).} \label{fig:figBeta}
\end{center}
\end{figure}

Fig. \ref{fig:figBeta} (top) shows the distribution of the slope
${\bf\beta}$ of the LDF function (see Eq. (\ref{ec:ldf})), for both
primaries and for two different spacings: 1500 m (\textit{left}) and
 750 m (\textit{right}), at a primary energy of 3.0 EeV,
i.e. at the lower limit of Auger full trigger efficiency. An
inspection of this top panel shows that ${\bf\beta}$ is quite
inappropriate for primary discrimination for the current Auger
spacing but it seems useful for the 750 m infill array.

Fig. \ref{fig:figBeta} (bottom) shows the $\bf{\beta}$ distributions
for  $10^{19}$ eV showers on the Auger surface array (\textit{left})
and $10^{18}$ eV showers on the 750 m infill array (\textit{right}).
This bottom panel shows that the 750 m infill has a similar
discrimination power as the full Auger array, but at an energy one
order of magnitude lower.

\begin{figure}[tbp]
\begin{center}
\includegraphics*[height=8cm,angle=-90]{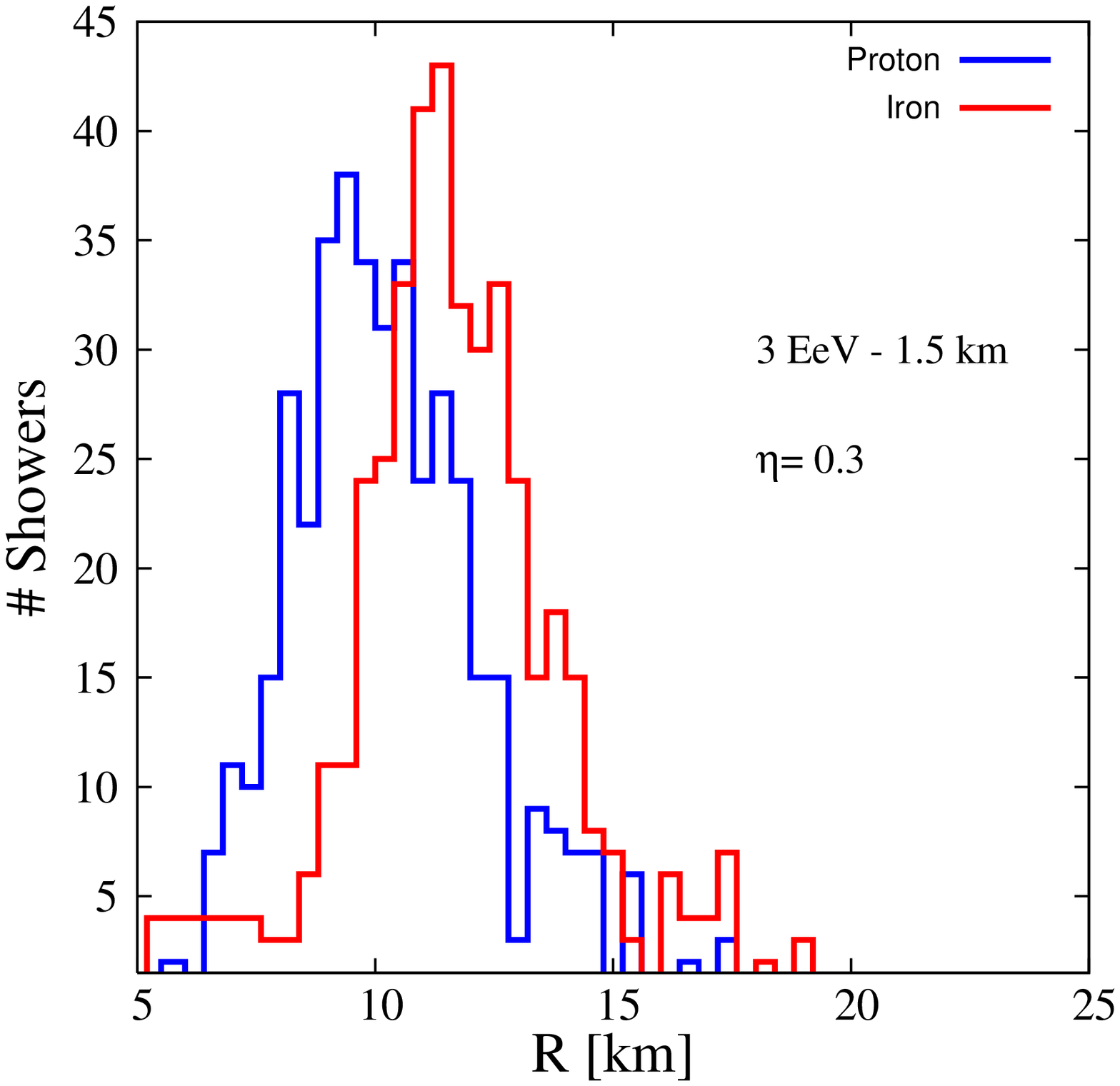}
\includegraphics*[height=8cm,angle=-90]{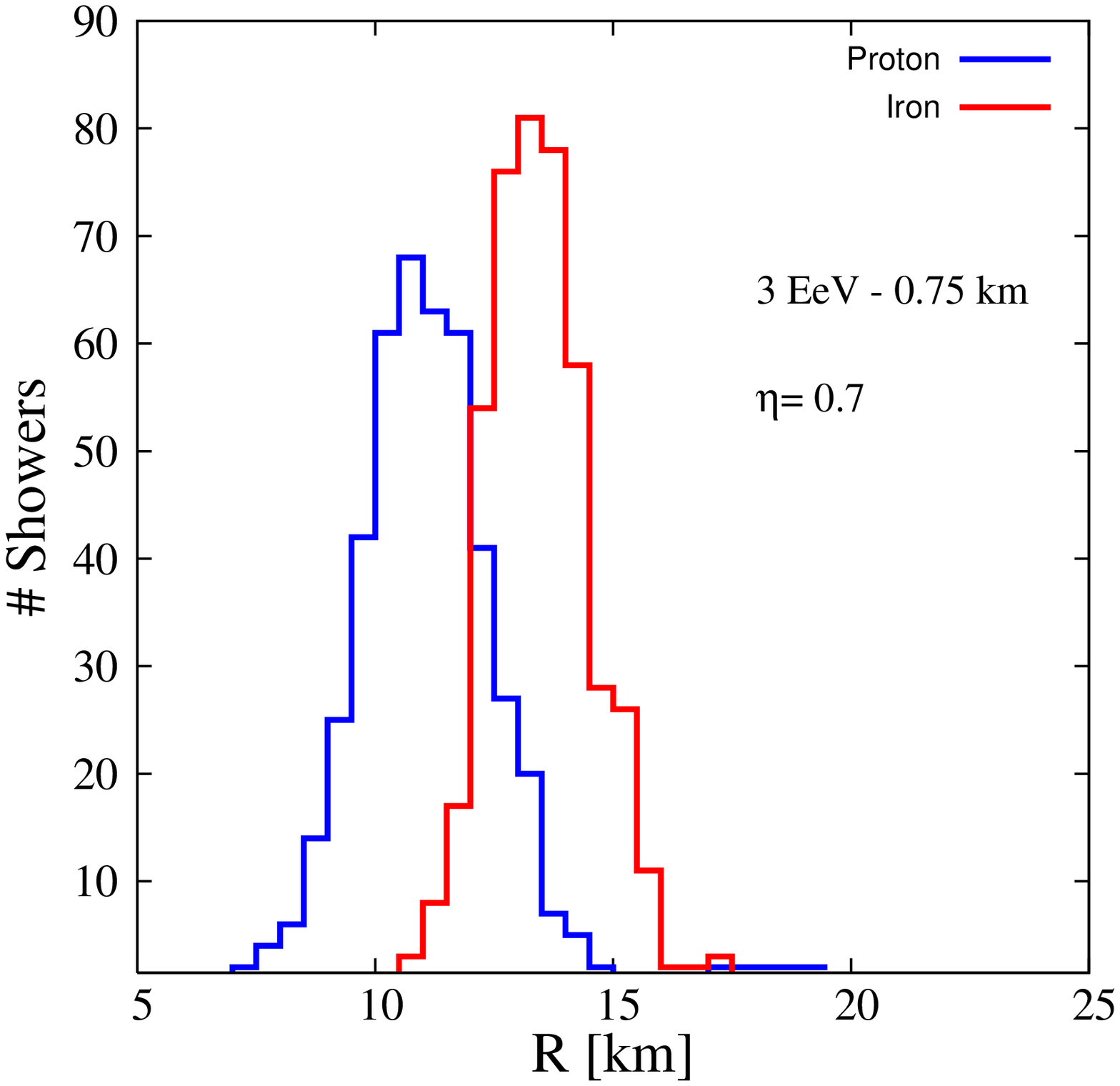}
\includegraphics*[height=8cm,angle=-90]{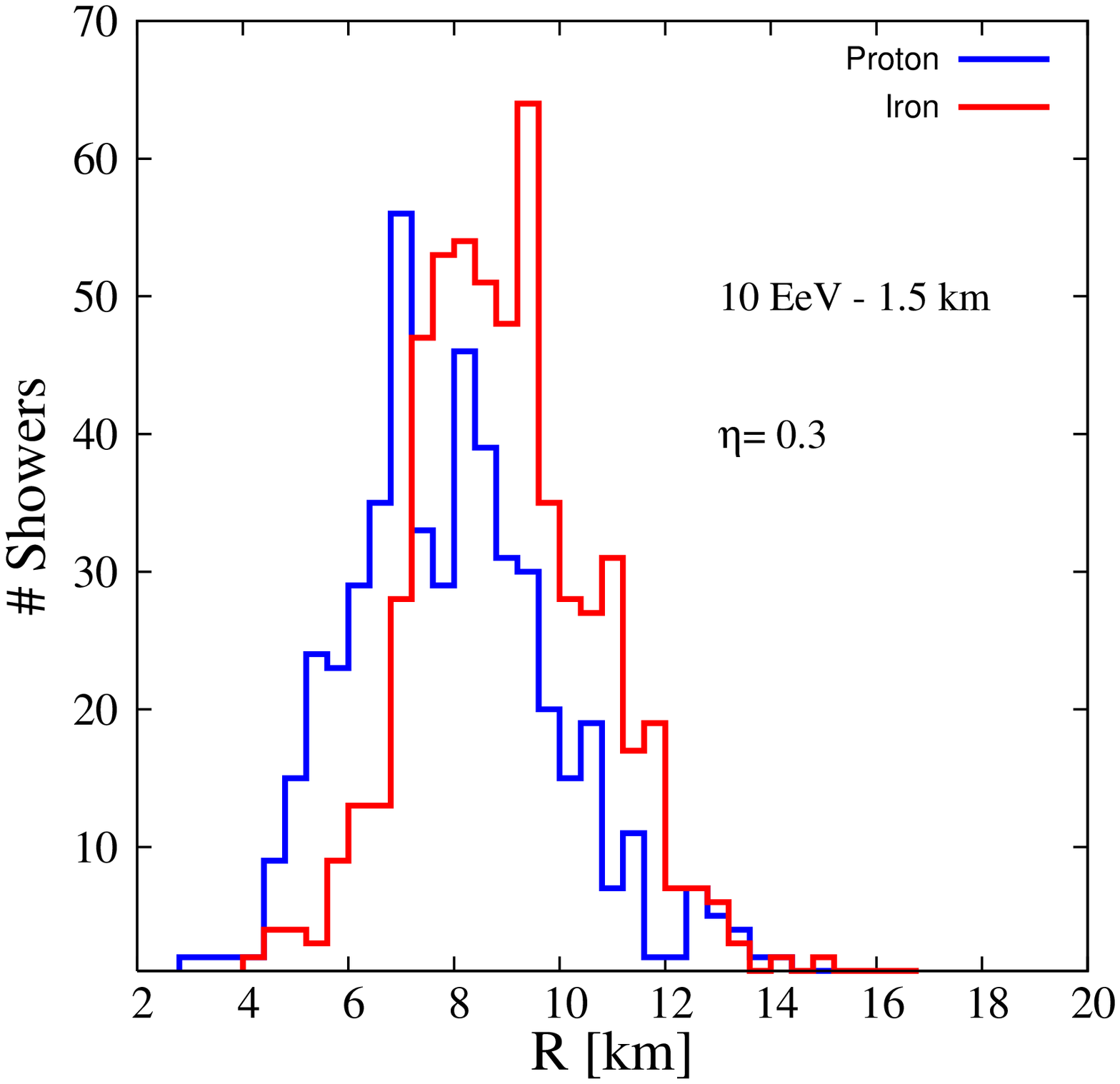}
\includegraphics*[height=8cm,angle=-90]{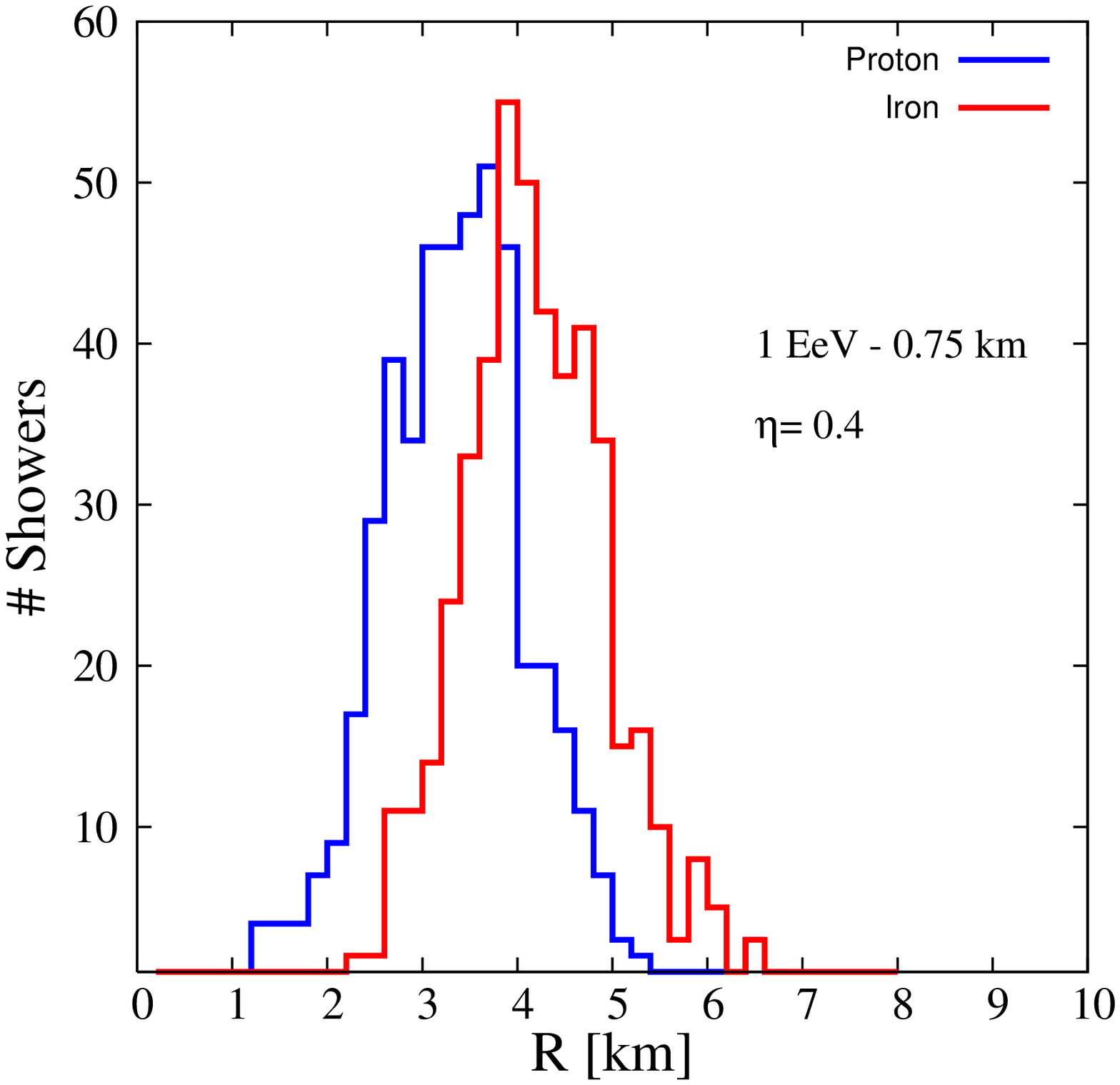}
\caption{\textbf{Top}: Distributions of the curvature radius of the
shower front for iron and proton at the limit of Auger acceptance,
i.e. $3\times10^{18}$ eV for Auger spacing (\textit{left}) and an
infill of 750 m (\textit{right}). \textbf{Bottom}: Distributions of
the curvature radius of the shower front for both primaries at
$10^{19}$ eV for a Auger 1500 m-array (\textit{left}) and at $10^{18}$
eV for a detector spacing of 750 m (\textit{right}).}
\label{fig:figRadCurv}
\end{center}
\end{figure}

Performing the same kind of analysis with the  radius of curvature
of the shower front, \textit{R}, yields similar results. In Fig.
\ref{fig:figRadCurv} (top) the distributions for $3\times10^{18}$ eV
are presented. Once again, at this energy the infill array helps to
statistically discriminate between proton and iron, whereas this is
seriously compromised for the Auger spacing. The bottom panel shows
that the power of discrimination at $10^{18}$ eV using the infill
information is slightly better than that of the present Auger
spacing for $10^{19}$ eV.

The improvement in these parameters gives a clear hint that
composition studies would greatly benefit from the information
provided by an infill array. Moreover, a multiparametric analysis as
suggested in \cite{Supa}, which weights in a large number of
variables, would allow a much more precise composition resolution.

\section{Conclusions}\label{sec:conclusion}

In this work we showed that enhancing the Pierre Auger Observatory
by increasing the density of surface detectors in a small area (20
$km^2$ of the total 3000 $km^2$) would bring considerable advantages
to the study of high energy cosmic ray physics, at a relatively low
cost (3$\%$ of the complete Auger SD). From the operational point of
view, such an enhancement could be smoothly integrated to the
existing surface array.

With an infill array much could be gained in shower reconstruction
at the lower limit of Auger energies: a better LDF fitting due to a
larger number of triggered detectors, a 2$\times$ and 4$\times$
improvement in arrival direction and core position resolution
respectively and a relative uncertainty of less than 15\% in
$S(600)$ determination.

Also, an infilled array with 750 m spacing is equivalent to 3
superimposed Auger-like arrays of 1500 m spacing. Single events can
then be reconstructed with 3 different subsets of detectors at Auger
spacings, allowing a direct study of fluctuations and uncertainties.
The data coming from an infill array would thus serve to check the
behavior of the original array and to validate further the
end-to-end simulation and reconstruction processes.

The extension of the detector's full efficiency interval down to
$\sim 10^{17}$ eV to englobe completely the second knee and the
ankle regions would be difficult to overestimate, allowing to test
in an unprecedented way competing models of Galactic and
extragalactic cosmic ray production and propagation. To this end,
better composition analyses could be performed since distinctions
among theoretical models largely depend on the primary type.
Furthermore, an additional  independent cross-check could be
achieved by overlapping with KASCADE-Grande in the lower energy
region.

\vspace{1.5 cm} \textbf{Acknowledgements}

We thank the Department of Mec\'{a}nica Computacional of Centro At\'{o}mico
Bariloche for providing us with the computing facilities for shower
simulations. This work is partially supported by CNPq and FAPESP
(Brasil) and CONICET, CNEA and IB (Argentina). We thank Dennis
Allard for providing us data for Fig. 1.



\newpage

\vspace{4 cm}

\vspace{4 cm}

\vspace{4 cm}

\vspace{4 cm}

\vspace{4 cm}

\vspace{4 cm}

\vspace{4 cm}

\vspace{4 cm}

\vspace{4 cm}

\vspace{4 cm}

\end{document}